# Magnetic Field Experiments Using a Phone

Glen D. Gillen and Katharina Gillen
Physics Department, California Polytechnic State University, San Luis Obispo, CA

December 5, 2021

**Abstract**

To make physics experiments more directly relevant to everyday life and help students realize how their smart phones or tablets can be used as sensors for scientific measurements, we designed two introductory physics experiments to measure vector magnetic fields. These experiments were initially developed for students working remotely in an online course setting. However, the experiments can also be used for in-person laboratory activities. The first activity uses the phone's three-dimensional sensors to determine the vector components and orientation of the background magnetic field at the students' locations. The resulting vector magnetic field is compared to the expected Earth's magnetic field for their location. The second experiment uses the phone sensors to measure the magnet field produced by the small magnet found in the speaker of common earbuds, or in-ear headphones. The students then carefully investigate the dependence of the magnitude of the field on the axial distance between the speaker magnet and the phone's sensors. Using a ring magnet approximation, advanced Excel curve-fitting methods are introduced to determine the radius of the magnet in the earbud and the distance between the magnet's location and the external surface of the earbud housing.

Experiment 1: Measuring the Earth's Magnetic Field Using a Phone

Experiment 2: Measuring the Magnetic Field of an Earbud Using a Phone

# Measuring the Earth's Magnetic Field
# Using a Phone

**Objective:** to determine the vector components, magnitude, and direction of the Earth's magnetic field using the sensors in a phone, or tablet, and compare your results to theory.

## I. Materials
- Device with magnetic field sensors (most any smartphone or tablet)
- Simple compass app[1]
- Phyphox app[2]
- Paper, pencil, and tape
- Word processing program to complete report

## II. Theory
The Earth's magnetic field is a three-dimensional vector field in and around the Earth as depicted in Figure 1(a). The vector magnetic field on the surface of the Earth varies depending upon your location. As with any vector, the Earth's magnetic field has both a magnitude and direction. The direction of the field for a specific location on the surface of the Earth is described by the **dip angle**, which is the angle between the local direction of the magnetic field vector and a line in the plane tangent to the surface (horizontal) which points North, or "horizontal north". The dip angles for 3 different locations are illustrated in Figure 1(b). The dashed line for each of the locations is north horizontal.

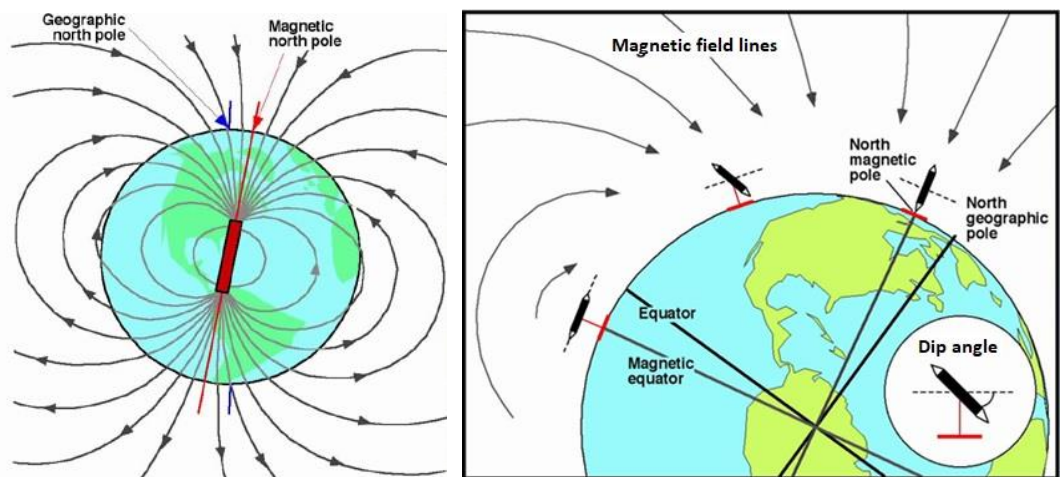

Figure 1. (a) Magnetic field lines of the Earth, and (b) the dip angle at three different locations on the surface. The dip angle is between the local direction of the magnetic field (3-D compass needle in the figure) and the dashed line (horizontal north).

For this experiment we will begin using a Cartesian coordinate system where the horizontal plane at your location will be the $x$-$y$ plane, and the $z$-direction will be the vertical direction. The exact direction of the $x$ and $y$ axes within the plane will depend upon the orientation of the sensors in your phone. (See discussion below.) The Earth's magnetic field at your location can now be written in component form as

Experiment 1: Earth's Magnetic Field

$$\vec{B}_E = B_x \hat{i} + B_y \hat{j} + B_z \hat{k}, \tag{Eq. 1}$$

or in magnitude-direction form as

$$\vec{B}_E = (B_E, \text{ dip angle}), \tag{Eq. 2}$$

where the magnitude of the Earth's magnetic field, $B_E$, can be calculated from the horizontal and vertical component magnitudes, $B_H$ and $B_V$, or

$$B_E = \sqrt{B_H^2 + B_V^2}. \tag{Eq. 3}$$

$B_H$ and $B_V$ are related to the chosen Cartesian coordinate system at your location by

$$B_H = \sqrt{B_x^2 + B_y^2}, \text{ and } B_V = B_z. \tag{Eq. 4}$$

The dip angle of Eq. 2, or $\phi_{dip}$, is the angle between the horizontal north direction and the magnetic field, and is defined in terms of magnetic field components as

$$\phi_{dip} = \tan^{-1}\left(\frac{B_V}{B_H}\right), \tag{Eq. 5}$$

and can also be expressed as

$$B_H = B_E \cos\phi_{dip}, \text{ and } B_V = B_E \sin\phi_{dip}. \tag{Eq. 6}$$

**Coordinate Axes and Phone Sensors**

The theory presented thus far has designated that the *z*-axis coincides with "vertical", or normal to the surface of the Earth, and the *x-y* plane coincides with "horizontal", or parallel to the surface of the earth at your location. The alignment of the *x* and *y* axes within the horizontal plane has intentionally been left undefined. The choice of this coordinate system is to match the axes of the three magnetic field sensors in the vast majority of phones.

Nearly all phones have the *z*-axis sensor aligned with the normal to the surface of the glass face of the phone. Additionally, nearly all phones have the *x* and *y*-axis sensors aligned perpendicularly to each other in the plane of the surface of the phone such that all three sensors are perpendicular to each other. However, many phones do not have the *x* and *y* sensors aligned with the perpendicular edges of the phone. In other words, if the phone is held horizontally and then rotated until the phone is pointing North, according to a compass app, that does not necessarily mean that the *y*-axis field sensor has a maximum magnitude and the *x*-axis field sensor has a reading of zero, or vice versa. In these cases, North has non-zero magnitudes for both the *x* and *y* field sensors.

Additionally, there are two coordinate systems: (1) the Earth's coordinate system designated by $\vec{B}_E$, $\vec{B}_H$, and $\phi_{dip}$, and (2) the *x*, *y*, and *z* axes of the sensors in your phone. The Earth's coordinate system is always constant; in other words, the vectors $\vec{B}_E$ and $\vec{B}_H$ always point in the same direction regardless of anything you are doing with your phone. However, the absolute direction of the *x*, *y*, and *z* axes *depends upon the orientation of your phone*, and there is no fixed relationship between this coordinate system and the Earth's coordinate system. Keep this in mind for Procedures B and C when you rotate your phone about a given axis, and think carefully about how your phone's "*x, y,* and *z* axes" relate to the fixed Earth's coordinate system throughout the rotation.

Experiment 1: Earth's Magnetic Field

**III. Procedure**

**A. Establishing the directions of North, East, $\vec{B}_x$, and $\vec{B}_y$**

1. Find an area of a table, or the floor, *away from any other sources of magnetic fields* (i.e., permanent magnets, laptops, tablets, computers, other technology or electronic equipment). If you are using the floor a non-carpeted floor would work best.
2. Tape, or secure, a sheet of paper such that it cannot slip.
3. Start your compass app[1].
4. Make sure your compass is calibrated. You can do this by simply moving your phone in the air in a figure 8 pattern while the compass app is running.
5. Set your phone down on the paper on your flat horizontal surface, and rotate the phone so the compass lines up exactly with "North".
6. **Draw a line on the paper** using the long side of your phone and label this line "N compass app". Be sure to **put an arrowhead on the line** to indicate the direction of north.
7. Repeat with East so that you have a line and directional arrow on your paper labeled "E compass app" indicating the direction of east. Make sure that the two lines cross each other.
8. Start the Phyphox app[2].
9. In the Raw Sensor section, start "Magnetometer".
10. Start recording by pressing the play button in the top right area of the screen.
11. Slowly rotate your phone until "Magnetometer x" is a maximum positive value.
12. Draw a line on the paper using the long side of your phone, indicate the direction on the line which points from the bottom of your phone towards the top. Label this line "B_x max". For best results, move (but do not rotate) your phone so that your line passes through the crossing of the N and E lines.
13. Repeat for "Magnetometer y" and draw a line, with a directional arrowhead, on the paper. Make sure that this line also passes through the crossing of the N and E lines and label this "B_y max".
14. Stop recording.

Question 1. In what direction (N, S, E, W, NNW, etc.) does $+\vec{B}_x$ point? (See the **Report** section at the end of this manual for instructions regarding your report and answering questions.)

Question 2. In what direction does $+\vec{B}_y$ point?

   **15. Take a picture of your paper and include this with your report.**

**B. Rotation about $z$ to determine $B_H$**
16. Set your phone down on your paper.
17. Clear the previous data from Part A by tapping the trashcan in the upper right corner.

18. Start recording a new data set and wait a few second before moving your phone to establish a steady-state signal. Slowly and smoothly rotate your phone 2 rotations with a constant angular speed, your rotation axis should be vertical and through the center of the phone.
19. Stop rotating your phone and wait a few seconds before pausing the data collection.
20. Hopefully, you have smooth "sine-wave" like curves with repeating equal maxima and minima in *x* and *y*, and a roughly constant value for *z*. If not, repeat until you do.
21. **Take a screenshot of your phone and include it with your report.** It should look similar to Fig. 2(a).

**Simple data analysis in Phyphox:**
- Tap on the plot of the dimension you want to analyze; for example, if you want to determine $B_{xmaxH}$, tap on the "Magnetometer x" plot.
- Tap on "Pick data".
    - To extract data from a single point, tap on a single point on your curve. A pop-up window will appear with the data for that point.
    - To extract the difference between two points, tap, hold, and slide your finger from the first to the second point. A pop-up window will appear with information between the two end points.

22. Analyze your data to determine $B_{xmaxH}$, the amplitude of the signal oscillation from the *x*-dimensional magnetic sensor of your phone for a rotation in the horizontal plane. (Reminder: you must neatly organize and show all your data and all your calculations in your report.)
    a. On the "Magnetometer x" graph use the two-point method for a line between a point at the very top of the signal oscillation to a point at the very bottom of the signal.
    b. Determine the amplitude of the oscillation, $B_{xmaxH}$
    c. Note that the vertical difference between a peak and a trough is *twice* the amplitude.
23. **Take a screen shot of your phone with your data analysis between two points and include it with your report.** This image should look similar to Fig. 2(b).

Experiment 1: Earth's Magnetic Field

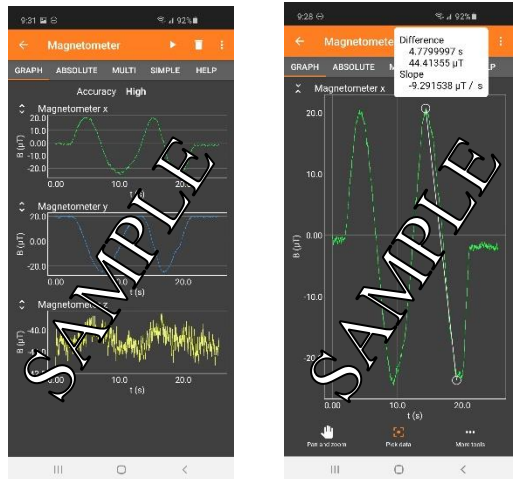

Figure 2. (a) Sample magnetometer data for rotation about *z*-axis, and (b) analysis of "Magnetometer x" signal.

24. Click the icon next to "Magnetometer x" to get back to the three-graph view.
25. Analyze your data from "Magnetometer y" to determine $B_{ymaxH}$, the amplitude of the signal oscillation from the *y*-dimensional magnetic sensor.
26. **Save a screenshot of the previous step and include this with your report.**

Question 3: What is the relationship between $B_{xmaxH}$ and $B_H$? Between $B_{ymaxH}$ and $B_H$? Explain clearly. Determine and clearly describe a method you can use to determine $B_H$.

27. Using your relationships between $B_{xmaxH}$, $B_{ymaxH}$, and $B_H$ determine a value for $B_H$.

## C. Rotation about *x* to determine $B_E$

28. Set your phone down on your paper and rotate it until $B_y$ is a maximum positive value.
29. Pick your phone straight up keeping its orientation in the *XY*-plane constant.
30. Clear the data to record a new data set.
31. Start recording and let it run for a few seconds without moving the phone to establish a constant initial signal.
32. Similar to steps 17-19, rotate your phone slowly and smoothly *about the x-axis* for two complete rotations. Repeat, if necessary, to get clean sine-wave-like plots for the *y* and *z* axes sensors.
33. **Save a screenshot of Magnetometer x, y, and z for rotation about *x*-axis for your report.**

Question 4: Analysis of which dimension(s) of your data (*x*, *y*, or *z*) would yield a value for $B_E$? Devise and clearly explain your method for extracting a value for $B_E$.

34. Determine your value for $B_E$.

**D. Calculating $\phi_{dip}$, and $B_V$**
   35. Use your values for $B_E$ and $B_H$ to calculate $\phi_{dip}$.
   36. Calculate the vertical component magnitude of the Earth's magnetic field, $B_V$.

Question 5: What is the Earth's magnetic field at your location? Express your answer in magnitude and direction format as in Eq. 2. If your lab partners are in a different location (say ≥ 100 miles) report their results and their location as well.

Question 6: To get another sense of the dip angle, start recording and rotate your phone in 3D until you reach the maximum possible $B_z$ reading. (It might be helpful to turn off any screen auto-rotate features before you do this.)
What is the relationship between this $B_z$ reading and your value for $B_E$? When your phone is held in this orientation what is the direction of the vector $\vec{B}_E$ relative to your phone? From where to where is the dip angle?

**E. Comparison of Experiment to Theory**
   37. Find your GPS coordinates in decimal format. There are a variety of ways to do this. One option is to go to https://gps-coordinates.org/my-location.php on your phone, laptop, tablet, etc., with GPS sensors. If you are using the "Just a Compass" android app click the three-dot menu icon in the top right and change the Coordinate System option to "Latitude & Longitude (DD)" and your GPS coordinates will be right below the compass on the main screen.
   38. Go to https://www.ngdc.noaa.gov/geomag/magfield.shtml, click on "Magnetic Field Calculator" and enter your GPS coordinates. (Note: if your longitude GPS coordinate value is negative then change your number to positive and click "West".)
   39. Record in your report the theoretical values for $B_H$ ("Horizontal Intensity"), $B_V$ ("Vertical Intensity"), $B_E$ ("Total Field"), and $\phi_{dip}$ ("Inclination").
   40. Calculate the % error between your measured values for each of these. (Remember to show all your work and calculations in your report.)

**App Footnotes**

[1] **Compass App:** For iOS phones there is a built-in compass app, 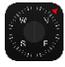, probably found in the utilities folder, 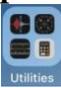. For Android phones, download a simple compass app from the Play Store. For example, "Just a Compass", 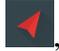, written by PixelProse.
[2] **Phyphox App:** Download from iOS App Store or the Google Play Store. The app is written by RWTH Aachen University and uses the 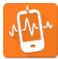 icon.

Experiment 1: Earth's Magnetic Field

**Report**
Create a neatly organized document as your report. Be sure to show all equations, data, work, calculations, etc. Include figure captions for any figures or screenshots. At a minimum, your report must have:
- Your name, and names of all lab partners
- Date and title of lab
- Picture of your paper for Part III.A. with all alignment lines, directions, labels, etc.
- Answers to Questions 1 – 5
- Screenshot of your phone for rotation about the $z$ axis
- Screenshots of your phone for analysis for $B_{xmaxH}$ and $B_{ymaxH}$
- All calculations for $B_H$
- Screenshots of your phone for rotation about $x$-axis, and analysis for $B_E$
- All initial equations, data, and calculations for $B_E$, $B_V$, and $\phi_{dip}$, and % error calculations
- Create a table with columns for: experimental value, theoretical value, and % error, and rows for each of the following: $B_H$, $B_V$, $B_E$, and $\phi_{dip}$

# Measuring the magnetic field of an earbud using a phone

**Objectives:**
- to accurately measure the magnetic field of the permanent magnet found in a speaker
- to investigate the field strength dependence upon axial distance
- to learn a curve-fitting procedure in Excel for user-defined non-linear functions
- to determine the approximate average radius of the speaker magnet
- to determine the physical distance between the speaker magnet and the phone's sensors when the earbud is placed in contact with the phone's screen

## I. Materials
- Device with magnetic field sensors (most any cell phone or tablet)
- Ruler with millimeters (or download a ruler app to use your phone as a ruler)
- Book with at least 1,000 pages; i.e., a physics textbook or other thick college textbook
- Earbuds or in-ear headphones[1]
- Phyphox app[2]
- Microsoft Excel with Solver Add-in[3], (Google Sheets Solver Add On[4])
- Word processing program to complete report

## II. Theory
At the heart of most any speaker is a permanent magnet and a coil of wire passing through a location where the field of the magnet is very strong. As current flows through the coil of wire (commonly called the "voice coil") the magnetic field exerts a force on the coil, which is attached to the cone of the speaker. If the electrical current in the coil oscillates with a certain frequency, the cone will move with that same frequency creating compressions and rarefactions in the air around the cone which, in turn, creates a sound wave.

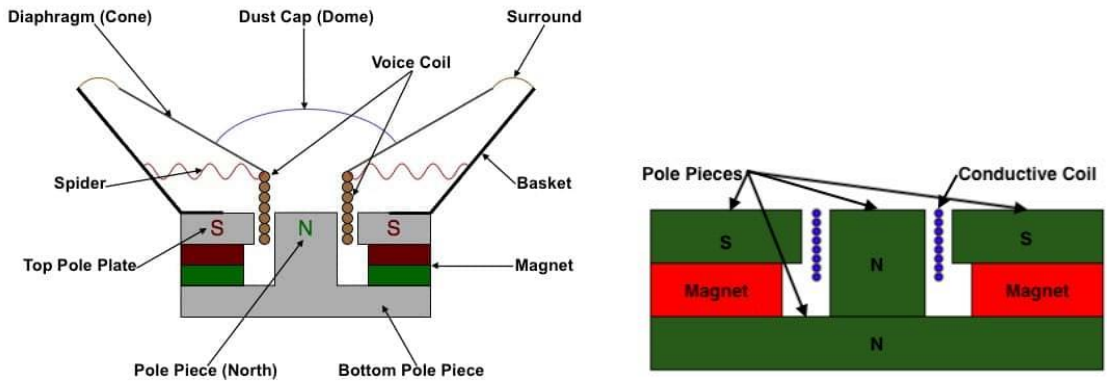

Figure 1. Components of a typical speaker.
[Image from https://mynewmicrophone.com/why-and-how-do-speakers-use-magnets-electromagnetism/]

Experiment 2: Earbud's Magnetic Field

The typical permanent magnet in a speaker has a complicated shape as illustrated in Fig. 1. However, it can be roughly approximated as an axial ring magnet, where the north and south poles are rings with a common axis and a common average radius as illustrated in Fig. 2.

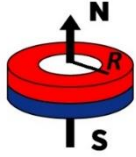

Figure 2. Axial ring magnet.

Using a cylindrical coordinate system with the z-axis passing through the center of the ring (colinear with the North-South line illustrated in Fig. 2.), a general equation for the axial component of the magnetic field along the *z*-axis can be expressed as

$$B_z = \frac{\mu_o}{2\pi} \frac{\mu}{\left(R^2 + z^2\right)^{3/2}},  \quad (Eq. 1)$$

where $\mu_0$ is the permeability of free space, $\mu$ is a constant for the strength of a specific magnet, *R* is the average radius of the ring, and *z* is the axial distance from the center of the ring.

### III. Procedure

**A. Distance Calibration**
For this activity we will need to be able to measure and change the distance between the earbud and your phone with high precision using only a traditional ruler. Ideally, we would be able to measure a change of distance less than 0.1 mm. However, a traditional ruler has markings only 1 mm apart; more than 10 times bigger than what we would like to be able to measure. The accuracy is further limited by the fact that the millimeter markings on a ruler can be up to a millimeter wide.
    Amazingly, your Physics 133 textbook not only contains the keys to understanding the world and universe around you, but can also be used as part of your physics experiments. (Note: if you have a digital copy of the textbook, instead of a physical copy you can substitute any thick book that has at least 1,000 pages or is at least 1-1/2 inches thick. You can perform this experiment at the library, if needed.) The thickness of a single sheet of paper can get us our desired measurement precision of less than 0.1 mm.

- Use a ruler to measure the thickness of a stack of pages close to the entire book; for example, measure from "Page 1" to at least "Page 1000". Make your measurement in millimeters. The higher the number of pages the higher your precision.
- Accuracy tip: Because the millimeter markings on a ruler are quite thick, adjust the number of pages such that the total thickness of the stack of pages lines up with the *exact middle* of the millimeter markings on your ruler. This might take some time and patience, but accuracy is our goal.
- Record the number of sheets of paper and thickness below and calculate the conversion factor. Note: **we are counting sheets of paper here, not page numbers**. The number of sheets of paper is half of the page number count and rounded up to an integer.

Experiment 2: Earbud's Magnetic Field

    Sheets of paper \_\_\_\_\_\_\_\_\_\_\_\_\_\_\_\_ Thickness \_\_\_\_\_\_\_\_\_\_\_\_ mm

    Conversion factor \_\_\_\_\_\_\_\_\_\_\_\_\_\_\_\_ $\frac{\text{meters}}{\text{sheet}}$

**Question 1:** What is the thickness of a single sheet of paper in your book?

**B. Initial magnetic field magnitude measurement, no sheets between earbud and phone**

First, let us define our coordinate system for the phone and measurements. The flat surface of the face of the phone is the *X-Y* plane, with the *x*-axis parallel to the bottom edge of the phone, and the *y*-axis parallel to the side of the phone. The *z*-axis is normal to the surface of the phone, or "straight up" from the screen.
- Make sure that your earbuds are unplugged, or turned off
- Start the Phyphox app
- Tap on "Magnetometer" under "Raw Sensors"
- Tap on the "Absolute" option
- Hold your earbud far away from your phone
- Click the data collect, or play, button towards the top right area of the screen
- Pause the data collection and determine the steady-state background magnetic field from your data.

**Question 2:** With your phone on the table and your earbud far away from your phone what is the magnitude of the background magnetic field?

- While recording data, hold the earbud on the screen with the speaker flat, or parallel, to the screen (as if the screen is listening to the earbud)
- Follow the procedure below to maximize the signal

**Procedure to maximize the signal**
- First identify the general location of the magnetic field sensors in the phone by sliding the earbud around on the screen until the "Absolute" reading gets big.
- Slowly slide the earbud in only the *y*-direction until the signal is a maximum
- At the location of the *y*-direction maximum, slowly slide the earbud along a line only in the *x*- dimension to find the new maximum.
- Repeat this process of horizontally sliding in the *x*-dimension at the vertical location of the previous *y*-maximum, go to the new *x*-maximum location, slide the earbud in a vertical line in the *y*-dimension which passes through the previous *x*-maximum. (You are now "walking" your way to a local two-dimensional maximum using a grid-pattern-style of searching.)
- **Did your field reading suddenly go to zero? If so, see Note [5] at the end.**

- Once you have found the location in the *X-Y* plane with the absolute maximum magnetic field reading, find the best angle along two rotation axes to try to maximize the measurement further.
    - Slightly and slowly rotate the earbud along an axis parallel to the *x*-axis until the signal is a maximum
    - Hold this angle along the *x*-axis
    - Slightly and slowly rotate the earbud now along an axis parallel to the *y*-axis until the signal is a maximum
- At this location, and angle, you should have the *absolute maximum* reading you can possibly get from the earbud.
- **Remember this location and angle to use as a starting point for maximizing the next measurement**
- Your phone screen should look like Fig. 3.
- **Save a screenshot of your phone display and include it with your report**.

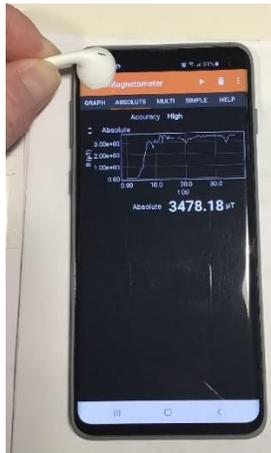

Fig. 3. Letting the data collection run while maximizing the signal in the *x*-dimension, *y*-dimension, angle along the *x*-axis, and angle along the *y*-axis.

- Pause the data collection.
- Record your maximum value, and units, for the magnetic field. You can tap to make the graph big, select "Pick data" at the bottom of the screen, and then tap on a data point that best represents the maximum value.

## C. Measuring the magnetic field as a function of distance along the *z*-axis
For this part, we are going to carefully, and accurately, increase the distance between the earbud speaker magnet and the magnetic field sensor in the phone.
- Place 10 sheets of paper (20 page numbers) between the phone screen and the earbud.
- In order to still be able to see the screen of your phone to make your measurement you can use the corner of the text-book pages between the earbud and the phone's screen. See Fig. 4.

Experiment 2: Earbud's Magnetic Field

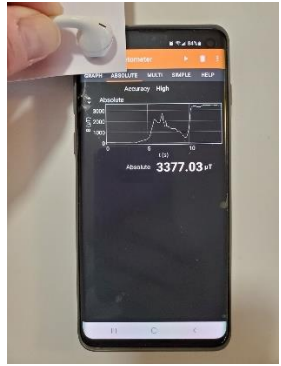

Figure 4. Using the corner of the textbook pages to still be able to see the screen of the phone.

- Repeat the procedure to maximize the signal.
- Record the maximum magnetic field measurement value.
- Construct a data table with the following columns. Use this specific formatting because you will need to add to the data table later for curve fitting and analysis.

|   | A | B | C |
|---|---|---|---|
| 1 |   |   | B (µT) |
| 2 | Sheets | z (m) | Experiment |
| 3 | 0 |   |   |
| 4 | 10 |   |   |
| 5 | ⋮ |   |   |

- Increase the number of sheets of paper by 10 each time, up to 80 sheets
- After 80 sheets, increase by 20 sheets each time (i.e., 80, 100, 120, etc.) up to 200 sheets
- After 200 sheets, increase by 100 sheets each time

## D. Calculating Chi-Squared for a user-defined function with initial approximate coefficients

Equation 1 can be written as

$$B(z) = \frac{A}{\left(R^2 + (z+z_o)^2\right)^{3/2}} + B_o \qquad \text{(Eq. 2)}$$

where $B$ is the magnetic field measurement as a function of the thickness of the paper between the earbud and the phone, $z$. The fit coefficients for the Chi-squared optimization are $A$, $R$, $z_o$, and $B_o$ where $R$ represents the average radius of the permanent magnet, $z_o$ represents the axial distance offset between the permanent magnet and the magnetic field sensors in the phone, and $B_o$ represents an offset due to any background magnetic fields. Even with no sheets of paper between the earbud and the surface of the phone there is a fixed distance, $z_o$, between the speaker

Experiment 2: Earbud's Magnetic Field

magnet and the phone sensor due to other physical components of the speaker, the earbud housing, the phone's screen, other components in the phone, etc.

- Add four more columns to your data table (columns D – G in the example below) for the Minimum Chi-Square Method of curve fitting routine.

|   | A | B | C | D | E | F | G |
|---|---|---|---|---|---|---|---|
| 1 |   |   | B (µT) | B (µT) | Relative |   |   |
| 2 | Sheets | z (m) | Experiment | Theory | Diff. squared | Value | Coef. |
| 3 | 0 |   |   |   |   | 0.005 | A (µTm$^3$) |
| 4 | 10 |   |   |   |   | 0.001 | R (m) |
| 5 | 20 |   |   |   |   | 0.005 | z_o (m) |
| 6 | 30 |   |   |   |   | 40 | B_o (µT) |
| 7 | 40 |   |   |   |   |   |   |
| 8 | ⋮ |   |   |   |   |   | Chi sq. |

- Use the starting values for your fit coefficients, as given in cells F3 to F6.
- Calculate the theoretical values for each axial distance, z (m), using Eq. (2). For example, in cell D3 type =$f$3/($f$4^2+(b3+$f$5)^2)^(3/2)+$f$6. Then copy that equation down column D for the rest of your axial distance values.
- Calculate the relative square of the differences between your experimental values and your theoretical values,

$$\text{relative difference squared} = \frac{(\text{experiment} - \text{theory})^2}{\text{theory}} \quad \text{(Eq. 3)}$$

for column E; i.e., cell E3 would be =(C3-D3)^2/D3. Copy this equation down column E for the rest of your axial distance values.

- Calculate the sum of the relative difference squared. This is called the "Chi-squared" value, or

$$\chi^2 = \sum_{i=1}^{N} \frac{(\text{exp.} - \text{theory})^2}{\text{theory}}, \quad \text{(Eq. 4)}$$

and put this value in cell F8 by typing =sum(E3:E##) into cell F8, where E## is the last data cell in your column E.

**E. Plotting *B* vs. *z* for both experimental and theoretical values**
- Make a complete graph of *B* vs. *z* using an XY scatter plot
- Set the series of your theoretical values to be displayed as a line, and your experimental values displayed as markers
- Adjust your values for *A* and $B_o$, if necessary, to get your experimental and theoretical plots roughly close. The Solver Add-In will fine-tune your coefficients. At this point if your two curves are within ~50% of each other that is plenty close enough for the Solver Add-in to take over.

**F. Using Excel's Solver Add-In for curve fitting**
Excel's solver add-in is not part of the default Excel installation, so you will need to install it. Follow the instructions at the end of this manual and install the add-in[3].
- Once you have finished installing the Solver Add-in click on it. You can find it under the "Data" tab on the main menu bar after clicking on "Data" it should be visible on the far right.
- A "Solver Parameters" window should pop up. The "Set Objective:" needs to be your Chi-squared value, or cell $F$8 in the previous example data table.
- Set the "To:" option to "Min".
- Set the "By changing variable cells:" option to your coefficient values, or $F$3:$F$6.
- Click "Solve"
- A "Solver Results" window should pop up. Be sure "Keep Solver Solutions" is checked, then click "OK" to accept the results

**Question 3:** What is your fit coefficient value (and units) for the radius of the permanent magnet in the speaker of your earbuds? Does this seem like a reasonable size?

**Question 4:** What is your fit coefficient for the absolute distance between the permanent magnet of your earbud, and the location of the magnetic field sensors inside your phone? Does this seem like a reasonable distance between them?

**Question 5:** We used paper as a spacer between the earbud and the phone. However, the equation we are using assumes there is vacuum in the space between the speaker and the sensor. What effect, if any, does the paper have on your measurements? Carefully explain your reasoning.
(Hint: you can safely assume that the magnetic properties of paper are similar to the magnetic properties of wood.)

Experiment 2: Earbud's Magnetic Field

**Report**
Create a neatly organized document as your report. Be sure to show all equations, data, work, calculations, etc. Include figure captions for any figures or screenshots. At a minimum, your report must have:
- Names of all lab partners
- Date and title of lab
- Screenshot of initial magnetic field maximization.
- Complete data table with your optimized fit coefficients
- Complete XY scatter graph of *B* vs. *z* with experimental results as markers and theoretical results as a line, graph title, axes labels with units, legend, etc.
- Answers to Questions 1 – 5

**Notes:**
[1] For this experiment we want to measure the field due to the small permanent magnet found in a small speaker such as an earbud or an in-ear headphone. The difference between the two is illustrated in the figure below. Earbuds rest outside your ear canal, and in-ear headphones are inserted into the ear canal. Since we want only the magnetic field from the speaker magnet (and not due to other electronic or magnetic sources) the simpler the earbud, or in-ear headphone, the better. Thus, wired earbuds or headphones might work better than wireless ones. If you have in-ear headphones with removable gel ear tips, remove the ear tips to get the speaker as close as possible to the phone's sensors.

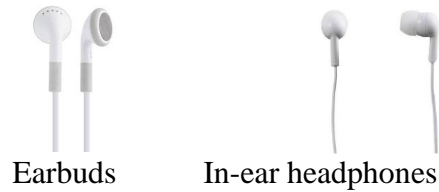

Earbuds    In-ear headphones

[2] **Phyphox App:** Download from iOS App Store or the Google Play Store. The app is written by RWTH Aachen University and uses the 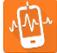 icon.

[3] **Excel Solver Add-in:**
- Excel's "Trendline" function is limited to only a few simple mathematical functions. However, with the Solver Add-in you can define your own fit function. The default Excel installation does not include this package so you must install it.
    - Open Excel and click "File" in the top left, then scroll down to the bottom and click "Options".
    - In the "Excel Options" window click on "Add-ins"
    - Click "Go…" at the bottom next to "Manage: Excel Add-ins"
    - Check "Solver Add-in" and click "OK"
    - Now on the main Excel toolbar under "Data" on the far right you should have an option for "Solver"

**[4] Google Sheets Solver Add On**
Google Sheets has a similar solver program. The instructions in this manual are for Excel's Solver Add-in. See [YouTube video](youtube.com/watch?v=JwYYIdmrmN8), youtube.com/watch?v=JwYYIdmrmN8, for information relating to the use of Google Sheet's Add On.

**[5] Magnetic field sensor turning off and Phyphox data suddenly going to zero?**
If you were recording the field strength and your value suddenly dropped to zero and stayed there, then your phone has deactivated the magnetic field sensors. Some models of phones have a self-protect mode and will shut off the magnetic field sensors if the magnetic fields get too high. (For most phones that do this the threshold seems to be ~3,000 to 4,000 µT.) If this is the case for you, then you will always have to keep a minimum number of pages between the earbud and the phone's screen. In other words, we need to place a known "spacer" between the earbud and the phone such that the sensor does not reach this cutoff threshold. Here are the steps to follow if this is the case for you:
- Reset the field sensor by fully closing the PhyPhox app
- Pick a small number of sheets of paper to put between the earbud and the phone. Start with perhaps ~20-30 sheets.
- Use only the corner of the pages between the earbud and the screen so that you can still see the screen.
- For finding the location of the sensors, instead of moving the earbud around the screen of the phone hold the earbud and the pages stationary and move the phone around underneath the pages as illustrated below.

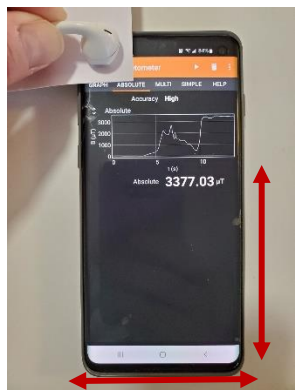